\documentclass[twocolumn,amsmath,amssymb,floatfix,pre]{revtex4}

\usepackage{graphicx}
\usepackage{bm}
\usepackage{natbib}

\newcommand{\Ca}{\mathrm{Ca}}
\newcommand{\ca}[1]{$\Ca=#1$}
\newcommand{\ph}[1]{$\phi=#1$}  
\newcommand{\tm}[1]{$t= #1$}  

\newcommand{\dbar}{$\bar{d}\; $}

\newcommand{\vp}{$\bar{v}_p \; $}

\begin{document}

\title{Segregation by membrane rigidity in flowing binary suspensions of elastic capsules}
\author{Amit Kumar}
\author{Michael D. Graham}%
\email{graham@engr.wisc.edu}
\affiliation{Department of Chemical and Biological Engineering, University of Wisconsin-Madison, Madison, Wisconsin 53706, USA}
\date{\today}

\begin{abstract}
\noindent Spatial segregation in the wall normal direction is investigated in suspensions containing a binary mixture of Neo-Hookean capsules subjected to pressure driven flow in a planar slit. The two
components of the binary mixture have unequal membrane rigidities. The problem 
is studied numerically using an accelerated implementation of the boundary integral
method. The effect of a variety of parameters was investigated, including the capillary number,
rigidity ratio between the two species, volume fraction, confinement ratio,
and the number fraction of the more floppy particle $X_f$ in the mixture. 
It was observed that in suspensions of pure species, the mean wall normal positions
of the stiff and the floppy particles are comparable. 
In mixtures, however, the stiff particles were found to be increasingly
displaced toward the walls with increasing $X_f$, while the floppy particles
were found to increasingly accumulate near the centerline with decreasing $X_f$. 
This segregation behavior was universally observed independent of the parameters.
The origin of this segregation is traced to the effect of the number fraction 
$X_f$ on the localization of the stiff and the floppy particles in the near
wall region -- the probability of escape of a stiff
particle from the near wall region to the interior is greatly 
reduced with increasing $X_f$, while the exact opposite trend is observed for
a floppy particle with decreasing $X_f$. Simple model
studies on heterogeneous pair collisions involving a stiff and a floppy
particle mechanistically explain the contrasting effect of $X_f$ on the near
wall localization of the two species. The key observation in these studies is
that the stiff particle experiences much larger cross-stream
displacement in heterogeneous collisions than the floppy particle.
A unified mechanism incorporating the wall-induced migration of deformable
particles away from the wall and the particle fluxes
associated with heterogeneous and homogeneous pair collisions is presented.
\end{abstract}

\maketitle

 \section{Introduction}
Flowing suspensions of mixtures of particles with different
rigidity are naturally encountered in blood flow.
Blood is primarily a suspension of red blood cells (RBCs) with small amounts of 
white blood cells (WBCs) and platelets. Normal RBCs are highly deformable,  and can thus pass easily through small capillaries, while leukocytes (WBCs)
and platelets are typically much stiffer. The leukocytes and platelets
are typically found in enhanced concentration near the walls -- this property is
commonly known as margination and is critical for the
physiological responses of inflammation and hemostasis \citep{eckstein87,munn09}. It is 
believed that the rigidity of the leukocytes and platelets 
play an important role in this margination phenomena. In contrast
to the benefits arising from the stiff nature of the leukocytes and platelets,
the stiffening of RBCs in various diseased states like malaria \citep{suwan04}
and sickle cell disease \citep{bunn97} leads to various deleterious effects.
A particular lethal complication of malaria
is cerebral malaria, in which infected RBCs are found to sequester in
the brain microcirculation \citep{kaul98}. This sequestration results from the
cytoadherence of the infected RBCs to the vessel walls of the 
post capillary venules \citep{kaul98}, and this effect is only expected to be 
exacerbated by the altered margination properties of the stiffened RBCs.
Recently, there has also been 
a particular emphasis on drug delivery with particles via the bloodstream. 
A  common target of such particles is the vascular
walls \citep{eniola10}, so in this context an ideal drug delivery particle must have an inherent
propensity to segregate toward the walls.  As such the design of an efficient
drug delivery particle (via shape, size, and deformability) is an interesting
problem in itself.

The segregation properties of leukocytes, platelets, and infected RBCs can 
also be employed for their separation or detection in biomimetic
microfluidic devices. For example, \citet{munn05} developed a network of rectangular microfluidic 
devices for the separation of leukocytes from whole blood. The process
of selectively removal of leukocytes is known as leukapheresis and 
has potential applications in decreasing the count of WBCs in patients 
with leukemia or for the removal of WBCs for transfusion \citep{sethu06}.
It was noted above that RBCs infected with the malaria parasite 
become stiff. \citet{lim10} employed this property to separate malaria-infected RBCs from the rest of the blood via the difference in 
margination properties based on stiffness. They showed that 
even a small difference in deformability is sufficient
to marginate the stiff infected RBCs, thereby enabling their separation.
A difference between that study and those on leukocytes is that
not only are leukocytes  stiffer than RBCs, they are also larger, which might play a role in their margination \citep{freund07}. The same is 
true for margination studies on platelets, which are stiffer and smaller
than RBCs \citep{zhao11,fogelson11}.

There have been several previous simulation-based efforts to investigate the mechanism(s) of the 
margination behavior of leukocytes and platelets. \citet{munn08} studied leukocyte margination and  concluded that the formation of stacks of 
discoidal platelets upstream of the leukocytes leads to the margination
behavior. \citet{freund07} saw a similar behavior in his study and 
suggested that the size difference between the RBCs and the leukocytes
could be playing a role. \citet{munn08} also studied suspensions of mixtures of particles
with different rigidity and found the stiffer particles to marginate toward the walls. They 
explained the phenomena based on single particle arguments, namely that the more
deformable particle has a greater tendency to migrate toward the centerline
and consequently displaces the stiffer particles. However, in a suspension, it is
expected that the particle-particle collisions will play an important,
perhaps dominant, role, and hence that single particle arguments may not be 
sufficient to provide a full explanation of the phenomenon.  \citet{fogelson11} studied numerically the margination behavior
of platelets in 2-D suspensions of blood. They described the phenomenon with a
drift-diffusion model, which has been used previously for interpreting experimental
results on the margination behavior of platelets \citep{eckstein91}. 
To obtain agreement between the  model and the detailed simulation results, 
\citeauthor{fogelson11} had to assume a drift velocity of the platelets
toward the wall when they are close to walls in the cell free
layer. They attributed this to the one-sided collision of platelets with RBCs, as no RBCs 
are present on the other side (on the side of the wall) to balance this drift.
There is, of course, nothing special about the platelets in the near wall region, 
as any particle in this region will undergo one-sided collisions,
and as a result get pushed toward the wall. This drift term, therefore, does not
provide any significant mechanistic insight into the segregation behavior, certainly
not for segregation in suspensions of equal sized particles.
In a similar study, using numerical simulations, \citet{zhao11} investigated the margination of 
platelets in blood flow. They concluded that the velocity fluctuations are responsible for this
behavior -- once the platelets marginate toward the walls, the small velocity fluctuations there
imply that they can't return back to flow. This argument, again, rests on the small size
of the platelets due to which they are influenced only by the local velocity
fluctuations, unlike the much larger RBCs. Consequently, the segregation 
between equal sized particles based on rigidity, if any, cannot be appropriately explained
by this mechanism.

The discussion above clearly establishes the importance of 
rigidity based flow induced segregation in mixtures of particles.
At the same time, it is also obvious that this 
phenomena is very poorly understood. In most previous works,
the key aspects of the proposed mechanisms appear to rest on the shape and the size discrepancy
between the mixture constituents -- the role of rigidity is not entirely clear and more broadly, no general formalism has been proposed for understanding the segregation phenomena.
In the present effort, we seek to delineate the effect of rigidity 
on flow induced segregation by focusing on binary mixtures of 
deformable capsules with unequal rigidities (i.e. stiff and floppy particles), but with identical size and spherical rest shape. The flow problem
involving suspensions of deformable capsules is investigated
numerically using an accelerated implementation of the boundary integral method.
The specific system geometry and bulk flow considered here is pressure driven 
flow in a planar slit.
The detailed simulation results indicate two key observations: (i) with increasing
fraction of floppy particles in the suspension, the stiff particles get 
increasingly localized in the near wall region, and (ii) with increasing
fraction of stiff particles in the suspension, the floppy particles are
increasingly depleted in the near wall region. 
As a result of this behavior, stiff particles segregate
in the near wall region in a suspension of primarily
floppy particles, while floppy particles segregate around the 
centerline in a suspension of primarily stiff particles.
The degree of localization of stiff and floppy particles
in the near wall region as a function of the fraction of the other
particle type is found to be greatly influenced by the cross-stream
displacement of that species in heterogenous pair collisions, i.e.
collisions involving a stiff and a floppy particle.  
In particular, in heterogenous pair collisions, the stiff particle undergoes
much larger cross-stream displacement than the floppy particle.
As a result, a stiff particle in the near wall region gets increasingly localized
as the fraction of floppy particles in the suspension increases, while the reverse
is true for a floppy particle in the near wall region as the fraction of stiff 
particles in the suspension increases. All these observations can be successfully
explained with a unified mechanism that incorporates the wall-induced migration
of deformable suspended particles away from the wall and the particle fluxes
associated with heterogeneous and homogeneous pair collisions.

The organization of this article is as follows. In Sec. (\ref{sec:form}) we 
formulate the problem and discuss the numerical solution procedure.
Next, in Sec. (\ref{sec:resA}), we present detailed results for one of the 
parameter sets explored in this work. This is followed by a detailed 
analysis of the segregation mechanism in Sec. (\ref{sec:mech}).
Concluding remarks are presented in Sec. (\ref{sec:conc}). 
Results for several other parameter sets 
are presented in the appendix.

\section{Problem formulation and implementation}\label{sec:form} 

\begin{figure}[!t]
\centering
\includegraphics[width=0.4\textwidth]{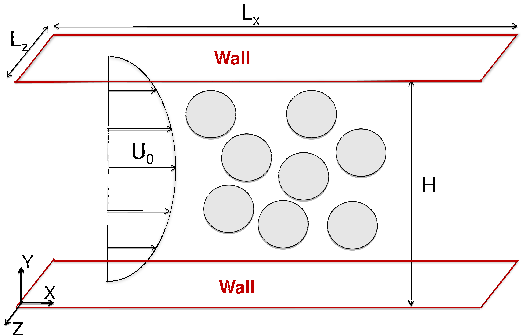}
\caption{(color online) A schematic of the problem set up.}\label{fig:geom}
\end{figure}

\subsection{Fluid velocity calculation: Boundary integral method}\label{sec:BIM}
We consider a suspension of fluid-filled deformable Neo-Hookean capsules between
two parallel plates as shown in Fig. (\ref{fig:geom}). Both the suspending fluid and the
fluid inside the capsules are assumed to be Newtonian and incompressible with the
same viscosity, $\mu$, i.e. the viscosity ratio $\lambda$ is unity.
We further assume that the Reynolds number for the problem is sufficiently small that the fluid motion
is governed by the Stokes equation. Under these assumptions, one
may write the fluid velocity $\mathbf{u}$ at any point $\mathbf{x}_0$ in the domain as
\citep{pozrikidis92}:
\begin{equation}\label{eq:BI}
u_j(\mathbf{x}_0) =   u_{j}^{\infty}(\mathbf{x}_0)  - \displaystyle \frac{1}{8\pi\mu} \displaystyle\sum_{n=1}^{N_p} \int_{S^n} \Delta f_i(\mathbf{x})\, G_{ji}(\mathbf{x}_0,\mathbf{x})\,dS(\mathbf{x}) 
\end{equation}
where $\mathbf{u}^{\infty}(\mathbf{x}_0)$ is the undisturbed pressure driven velocity
at  point $\mathbf{x}_0$, $S^n$ denotes the surface of particle $n$,
$\Delta \mathbf{f}(\mathbf{x})$ is the hydrodynamic traction jump across the interface,
while $\mathbf{G}$ is the Green's function for the Stokes equation in the geometry of interest. Note that the sum in the above expression
is over all the $N_p$ particles in the system. A crucial aspect of the above
formulation is that the Green's function $\mathbf{G}$ is taken to satisfy
the boundary conditions imposed at the system boundaries, so the integrals above
only involve the internal (interfacial) boundaries; if the Green's function 
for any other geometry is employed (e.g. free-space), additional integrals
over the domain boundaries will arise in Eq. (\ref{eq:BI}).
For the present case of a slit geometry, we have periodic boundary conditions
in $x$ and $z$ directions with spatial periods $L_x$ and $L_z$, respectively (Fig. \ref{fig:geom}).
In the $y$ direction, we have no slip velocity boundary conditions at the two walls
at $y=0$ and $y=H$ (Fig. \ref{fig:geom}).

Traditional implementations of the boundary integral method typically
have a scaling of at least $O(N^2)$, where $N$ is proportional
to the product of number of particles $N_p$ and the number of elements  $N_\Delta$
employed to discretize the particle surface. In the present effort,
we employ an accelerated implementation of the boundary integral
method with a computational cost of $O(N\log N)$ for the slit geometry \citep{kumar_jcp}.
The acceleration in our implementation is provided by the use of the General 
Geometry Ewald like method (GGEM) \citep{ortiz07}, which essentially provides fast 
solution for the velocity and the pressure fields driven by a collection
of point forces in an arbitrary geometry -- finding these velocity and pressure 
fields is at the heart of the solution procedure of the boundary integral method \citep{kumar_jcp}.
The key aspect of the GGEM methodology is to split a Dirac-delta force density 
into a smooth quasi-Gaussian \textit{global} density $\rho_g(\mathbf{r})$
and a second \textit{local} density $\rho_l(\mathbf{r})$;
these are respectively given by the following expressions:
\begin{subequations}\label{eq:rho_l_g}
\begin{equation}
\rho_g(\mathbf{r}) = \frac{\alpha^3}{\pi^{3/2}}e^{-\alpha^2 r^2} \left(\frac{5}{2}-\alpha^2 r^2 \right),
\end{equation}
\begin{equation}
\rho_l(\mathbf{r}) = \delta(\mathbf{r})-\rho_g(\mathbf{r}),
\end{equation}
\end{subequations}
where $\alpha^{-1}$ represents a length scale over which the delta-function
density $\delta(\mathbf{r})$ has been smeared using the quasi-Gaussian form above, 
while $\mathbf{r}$ is position vector relative to the pole of the singularity.
It is important to emphasize that the total density remains a $\delta$-function, i.e.
$\rho_g(\mathbf{r}) + \rho_l(\mathbf{r}) = \delta(\mathbf{r})$.
The solution associated with a local density, which is known analytically,
is short ranged, and is neglected beyond a length scale of $O(\alpha^{-1})$
from its pole. We note that the local solution is obtained assuming
free-space boundary conditions.  The solution associated
with the global density is numerically computed, while ensuring that the boundary conditions
associated with the overall problem are satisfied \citep{ortiz07,kumar_jcp}.
Complete details of this method can be found in \citep{kumar_jcp}; in short, for the slit geometry of interest here we have implemented a spectral Galerkin 
scheme with Fourier series representation in the periodic $x$ and $z$ directions
and Chebyshev polynomial series representation in the $y$ direction; the number of
corresponding modes employed in $x$, $y$, and $z$ directions are denoted by 
$N_x$, $N_y$, and $N_z$, respectively. An important parameter controlling the error 
of the GGEM solution is $\alpha h_m$, where $h_m$ is the mean mesh spacing associated 
with the global solution procedure (e.g. $h_m=L_x/N_x$). Based on extensive tests
in \citep{kumar_jcp}, we have set $\alpha h_m = 0.5$ in this study with the mean mesh spacing 
being equal in all three directions. The value of $\alpha$ was held constant in all
cases: $\alpha=4/a$, where $a$ is the radius of the spherical capsules at rest.

\subsection{Surface discretization and membrane mechanics}\label{sec:memb}
The surface of each of the particles is discretized into $N_{\Delta}$
triangular elements with linear basis functions employed over each element.
In the present study, we took $N_{\Delta}=320$. As will be seen later,
 long simulations are required to reach a steady
state, and this mesh resolution keeps the computational time
requirement manageable. All simulations reported here were performed on a single processor.

The computation of the boundary integral in Eq. (\ref{eq:BI}), requires 
the knowledge of the hydrodynamic traction jump across the interface ($\Delta \mathbf{f}$).
This jump in traction is obtained from the membrane equilibrium condition, which requires
that the hydrodynamic forces on any infinitesimal area of the membrane be 
balanced by the elastic forces in the membrane \citep{pozrikidis92}. Hence $\Delta \mathbf{f}$
can be obtained from the knowledge of elastic stresses. The 
capsule membrane is modeled here as a neo-Hookean membrane with a shear modulus $G_s$
\citep{Barthes-BieselDD02}. The elastic forces in the membrane are then computed using the 
approach of \citet{CharrierSW89}, which is based on the principle of virtual work.
Implementation details of this approach can be found in \citep{kumar_jcp,pratik10}.

\subsection{Physical parameters}\label{sec:prm}
We detail here the various physical parameters relevant for this study.
To begin, we note that the rest shape of all capsules
is spherical with radius $a$. This radius $a$ will be taken as the unit
of length throughout this work. Next, we introduce the undisturbed
pressure driven flow, whose velocity in the flow direction $x$
is given by:
\begin{equation}
u = 4 U_0 \; \frac{y}{H} \; \left(1- \frac{y}{H}\right),
\end{equation}
where $U_0$ is the centerline velocity. Based on this flow
profile, we obtain a wall shear rate of $\dot{\gamma}_w=4U_0/H$.
Time in this work will be presented in units of $\dot{\gamma}_w^{-1}$.
The shear modulus of the capsule will be expressed by the non-dimensional
capillary number  $\Ca=\mu \dot{\gamma}_w a/G_s$, where $\mu$ is the 
viscosity of the suspending fluid. The capillary number can be 
viewed as the ratio of viscous and elastic stresses on the capsule.
Since we study suspensions of binary mixtures of capsules with different rigidity,
we remark that in a given flow, the values of $\Ca$ of the different particles are not equal: 
particles with the lower $G_s$ have a higher $\Ca$ and are termed
floppy particles, while particles with the higher $G_s$ have a lower
$\Ca$ and are termed stiff particles. The ratio of the rigidities ($G_s$) of 
the stiff particle and the floppy particle will be denoted by $R$, with $R \geq 1$.
Another important parameter characterizing a binary mixture is the number fraction of one of  
species in the suspension -- in this work, we employ the number fraction of the
floppy particles $X_f$. The number fraction of the stiff particles can be
obtained from $X_f$ as: $X_s=1-X_f$. The suspension as a whole is 
characterized by its volume fraction $\phi$, so for example $X_{f}\phi$ is the overall volume fraction of floppy particles.
 
In the present work, we have neglected bending resistance of the capsule membranes. In such a case, compressive
stresses in the membrane can lead to membrane buckling at both at low and high $\Ca$ \citep{lac07}. 
In the buckled state, the numerical result is likely spurious and does not
represent true physics. A  simple way to alleviate this problem
is to introduce an isotropic prestress in the membrane, which can also
arise naturally in experiments due to osmotic effects \citep{risso03}.
As a result of the prestress, the rest radius of the capsule $a$
is larger than the unstressed radius $a_0$. The degree of inflation
can be characterized by the inflation ratio $beta = a/a_0-1$. In
pair collision studies, inflation ratios between $0.05-0.1$ are 
typically required to prevent buckling depending on the $\Ca$, though,
at higher $\Ca$ even these inflation ratios are insufficient \citep{lac07}. 
In this work, we set the inflation ratio to $\beta=0.1$. We did not
observe any buckling instability in our simulations. The effect
of preinflation on the cross-stream displacement in pair collisions
is usually weak \citep{lac07}. We emphasize that the inflated radius $a$ is taken
as the relevant radius of the capsule; for example, $\Ca$ is defined
based on the inflated radius $a$.

\begin{table*}
\caption{Parameter specification in various simulation sets.}
\begin{center}\label{tb:runs}
\begin{tabular}{|c|p{1cm}|p{1cm}|p{1cm}|p{1cm}|p{1cm}| p{1cm}|p{1cm}|}
\hline
Set & $\Ca_1$ & $\Ca_2$ & $R$ & $X_f$ & $\phi$ & $2a/H$  & $N_p$\\
\hline
A  & 0.2 & 0.5 & 2.5 & 0 -- 1 & 0.2 & 0.197 & 50 \\ 
B  & 0.2 & 0.5 & 2.5 &  0 -- 1 & 0.2 & 0.3  & 50 \\
C  & 0.2 & 0.5 & 2.5 & 0 -- 1 & 0.12 & 0.197 & 30  \\
D  & 0.1 & 0.6 & 6.0 & 0 -- 1 & 0.12 & 0.197 & 30  \\
\hline
\end{tabular}
\end{center}
\end{table*}

Having defined all the important parameters, we next tabulate
the set of simulations performed in this study with  their specific choices
of parameters (Table \ref{tb:runs}). As can be seen, we have
performed four main sets of simulations, denoted  A, B, C and
D in the table. These sets of simulations address the 
effect of the number fraction of the floppy particle $X_f$ 
on the segregation behavior in combination with several 
other  parameters including $\phi$, $\Ca$, $R$ and 
the confinement ratio $2a/H$.

\subsection{Simulation details}
We next outline the major steps involved in the solution procedure. The first
step involves the computation of the hydrodynamic traction jump $\Delta \mathbf{f}$ --
this is computed using the approach outlined above in Sec. (\ref{sec:memb}).
Following this, we compute the velocity at all surface element nodes employing 
a discretized version of the boundary integral Eq. (\ref{eq:BI}) \citep{kumar_jcp}.
The surface element nodes are then advanced in time using the second order
explicit Adams-Bashforth method. The time step $\Delta t$ is set adaptively to satisfy:
$\Delta t = 0.5\, Ca \, h$, where $h$ is the minimum 
node-to-node distance (the nearest nodes do not have to be on the same particle).
Recall that time is expressed in units of $\dot{\gamma}_w^{-1}$, while
distances are expressed in units of $a$.  Although the above choice of time step $\Delta t$
is sufficiently small in general,  occasional particle-particle overlaps can occur.
In principle the time step can be reduced further until overlaps are eliminated,
but the computational cost can become exorbitant. Alternatively, 
the overlaps can be prevented by employing a short range repulsive
force, though this is not very effective in three dimensional
simulations \citep{freund10}. The best approach appears to be an overlap correction
in an auxiliary step \citep{freund10}. This approach is also frequently used
in suspensions of rigid particles \citep{meng08,kumar10}. In the present work,
like in suspensions of rigid particles \citep{meng08}, we not only correct the
overlaps, but also maintain a minimum gap ($\delta_{min}$) between the surfaces of two particles.
In the present case, we set the minimum gap  to a small value of one percent of the (rest) particle 
diameter, i.e. $\delta_{min}=0.02a$. Maintaining the minimum gap is also justified 
in the absence of adaptive meshing, as that will be necessary 
to accurately resolve the flow dynamics in the small gaps.
No adaptive meshing was employed in this work. The overlap
correction procedure involves moving a pair of overlapping particles 
apart along their line of centers like a rigid particle until the 
minimum gap requirement is satisfied. Translating  the capsules like
a rigid particle in this auxiliary step has the benefit that the
shapes of the particles remain unchanged, as is the orientation of
the particles with respect to the flow. 
In general, multiple steps of the correction procedure are required,
as the correction of overlap between one pair could result in other 
overlaps \citep{meng08}. This overlap correction step involves 
minimal displacement of the particles, on the order of $\delta_{min}$. 
Given that the maximum volume fraction studied in this work
is $\phi=0.2$, this procedure was rarely required  --
only a couple of minimum gap violations were observed per time unit.
Time integrations were performed for sufficiently long times (hundreds of time units) to allow development of an statistical steady state in the particle distribution.

\section{Results}\label{sec:resA}

\begin{figure}[!t]
\centering
\includegraphics[width=0.25\textwidth]{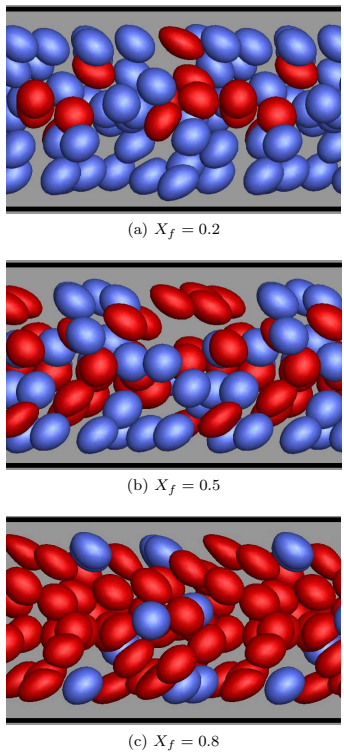}
\caption{(color online) Simulation snapshots in the shear plane ($x-y$) for different values of $X_f$.
The snapshots show approximately 1.5 times the length of the simulation box
in the $x$ direction. The stiffer particles are light (blue online), while the floppy
	particles are dark (red online).}\label{fig:50snap} 
\end{figure}

We begin by discussing the results for simulation set A of table (\ref{tb:runs}).
In this set of runs, the stiffer particle has a capillary number of $\Ca=0.2$,
while the floppy particle has a capillary number of $\Ca=0.5$, such that  
the rigidity ratio $R=2.5$. The confinement
ratio for this set of runs was fixed at $2a/H=0.197$. 
The number fraction of floppy particles in the mixture ($X_f$)
was varied between 0 and 1, with $X_f=0$ corresponding to a pure suspension
of stiff particles ($\Ca=0.2$) and $X_f=1$  corresponding to 
a pure suspension of floppy particles ($\Ca=0.5$). Several simulation snapshots
for different values of $X_f$ are shown in Fig. (\ref{fig:50snap}) to help
in visualizing the problem set up. We also remark that, throughout this article,
we will often use the letter `S' to denote stiff particles, and the letter `F' to
denote floppy particles.

\begin{figure}[!t]
\centering
\includegraphics[width=0.35\textwidth]{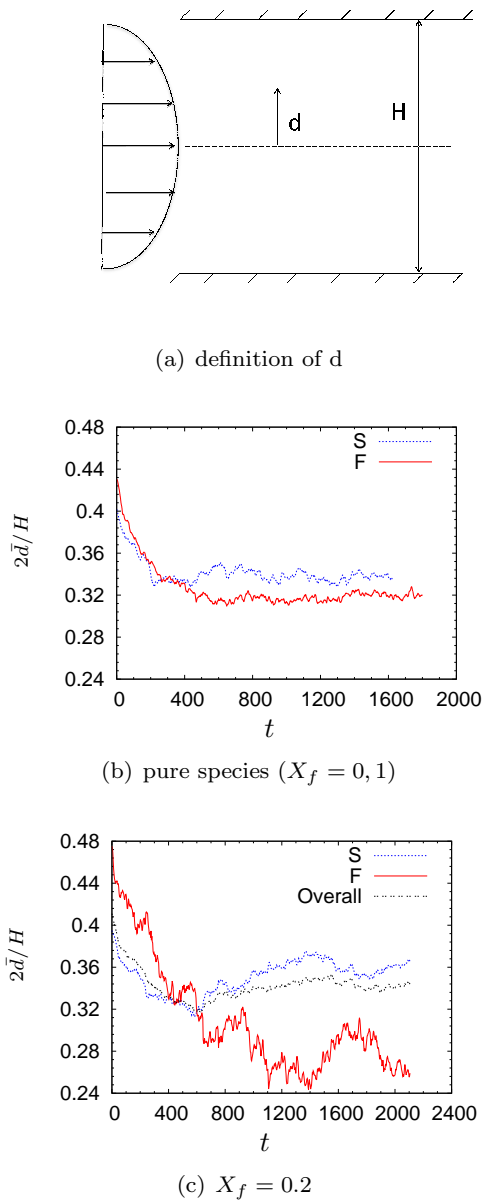}
\caption{(color online) (a) Definition of d: it gives the absolute distance from the centerline.
(b) \& (c)  Time evolution of the mean absolute distance
of a species from the centerline $\bar{d}$ non-dimensionalized by half the channel width $H/2$.
In the plots `S' refers to the stiff particles ($\Ca=0.2$), while `F' refers
to the floppy particles ($\Ca=0.5$). In (b) results are shown for suspensions
of pure species ($X_f=0$ \& $X_f=1$), while in (c)  results are shown for a suspension
with $X_f=0.2$.}\label{fig:ycm_t}
\end{figure}

\begin{figure}[!t]
\centering
\includegraphics[width=0.35\textwidth]{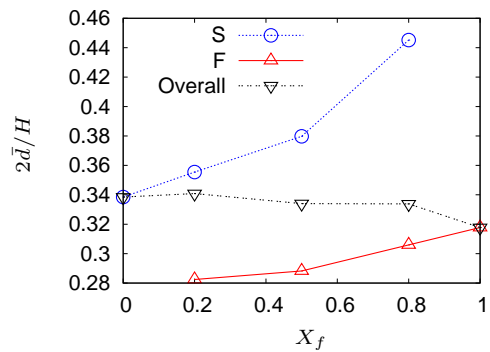}
\caption{(color online) Steady state \dbar as a function of $X_f$ for the \ca{0.2} and \ca{0.5} particles.
	Also shown is the steady state \dbar of the ``overall'' suspension.}\label{fig:y_avg}
\end{figure}

\subsection{Particle distribution in the wall normal direction}\label{sec:dist}
The first suspension property we examine is the distribution of particles in
the wall normal direction $y$. To characterize this distribution
of particles, we will employ three complementary measures. In the first measure,
we compute the mean absolute distance of a species from the channel centerline, denoted by $\bar{d}$;
see Fig. (\ref{fig:ycm_t}a). This simple measure, though, does not provide detailed information about the actual distribution 
of particles along the wall normal direction. In order to quantify this distribution, therefore, we 
will use two additional measures: the first is $\hat{\phi}(y)$, which gives the 
volume fraction of a given species as a function of $y$-coordinate normalized
by the mean volume fraction of that species, 
while the second measure is $\hat{n}(y)$, which gives the number density of the center of mass
of a given species as a function of the $y$-coordinate normalized by the mean
number density $n$ of that species. If the distribution were uniform, then both of these measures
would have a value of unity throughout. Note that a subscript `s' or `f' may be used with $\hat{\phi}$ or $\hat{n}$
for specifically referring to the stiff or the floppy particles, respectively; the same convention
is followed later for denoting other quantities. We now briefly describe the procedure
to compute $\hat{n}(y)$; a  similar procedure is employed to compute $\hat{\phi}(y)$.
For this calculation, the channel height is first divided into bins and then the  particles are assigned
to  bins based on their center of mass coordinates. The number of particles in each of the bins is then 
normalized by the value expected in that bin based on a uniform distribution in the 
wall normal direction.

With the different measures for characterizing the particle distribution specified,
we turn to the results for these measures. We show the time evolution of $\bar{d}$ for the case
of pure species (i.e. $X_f=0$ \& $X_f=1$) in Fig. (\ref{fig:ycm_t}b) and for a mixture
with $X_f=0.2$ in Fig. (\ref{fig:ycm_t}c). It can be seen that in all cases  $\bar{d}$
initially decreases with time, indicating that the particles are drifting toward the centerline.
This behavior is not only consistent with results in suspensions of deformable particles 
subjected to pressure driven flows \citep{munn08,bagchi09}, but also with results 
in suspensions of rigid particles \citep{phillips92,nott94}. 
However, as we shall see shortly, the behavior in suspensions
of particle \textit{mixtures} is much richer. In suspensions of pure species
(Fig. \ref{fig:ycm_t}b), we find that $\bar{d}$ reaches an apparent steady state
relatively quickly in about \tm{500}. In addition, the value of \dbar for stiffer particles ($\Ca=0.2$)
is found to be slightly higher than for floppy particles ($\Ca=0.5$), though the difference is not
large. 

Next, we consider results for the mixture with $X_f=0.2$ 
(Fig. \ref{fig:ycm_t}c). In this figure, we show \dbar for the stiff particles (\ca{0.2})
and the floppy particles (\ca{0.2}) individually; in addition, \dbar for the overall suspension is also shown. First
of all, one can note that the \dbar for \ca{0.2} particles and \dbar for \ca{0.5} particles 
are different at $t=0$. This is due to the finite system size with the total number of
particles being $N_p=50$, and the average being performed over 2 realizations;
if large number of realizations are used, then \dbar of both the species will be the same,
though this is not expected to affect the steady state distribution.
Returning to the figure, we observe that initially both the
stiff and the floppy particles drift toward the centerline. However, beyond approximately
\tm{600}, in what appears to be a slowly evolving process, the stiff
particles are displaced toward the walls by the floppy particles.
The \dbar of the overall suspension, though, shows  much weaker 
evolution for $t >800$, thereby confirming that this phase of
the simulation mostly involves rearrangement of positions between the two particle types.
\citet{munn08} saw a similar behavior in their simulations on mixtures of deformable particles,
which led them to the same conclusion as here. The mechanism underlying this 
observation is described in Sec. (\ref{sec:mech}).

We next summarize the results for the steady state 
\dbar of both the species at all values of $X_f$ studied here (Fig. \ref{fig:y_avg}).
Note that the steady state \dbar was obtained by averaging over the last 
1000 time (strain) units of the simulation. Two trends are immediately obvious in this plot. First
is that \dbar for both the stiff and the floppy particles increase with increasing $X_f$,
though \dbar for the whole suspension actually decreases with increasing $X_f$. The latter
is expected as with increasing $X_f$, the suspension, on average, becomes more deformable
and hence has a lower \dbar. An alternative interpretation of the above observation is that
with increasing fraction of the floppy particles, the stiffer particles get displaced toward
the wall, while  with increasing fraction of the stiffer particles, the floppy particles get
displaced toward the centerline. The second observation is on the rate of increase or decrease
in \dbar of a given species with changing fraction of the other species: the rate 
of increase in \dbar of stiffer particles is  more rapid with increasing $X_f$
than is the rate of decrease in \dbar of floppy particles with decreasing $X_f$. This
points toward a varying degree of segregation with changing $X_f$, which we will 
quantify shortly with an appropriate measure.

\begin{figure}[!t]
\centering
\includegraphics[width=0.35\textwidth]{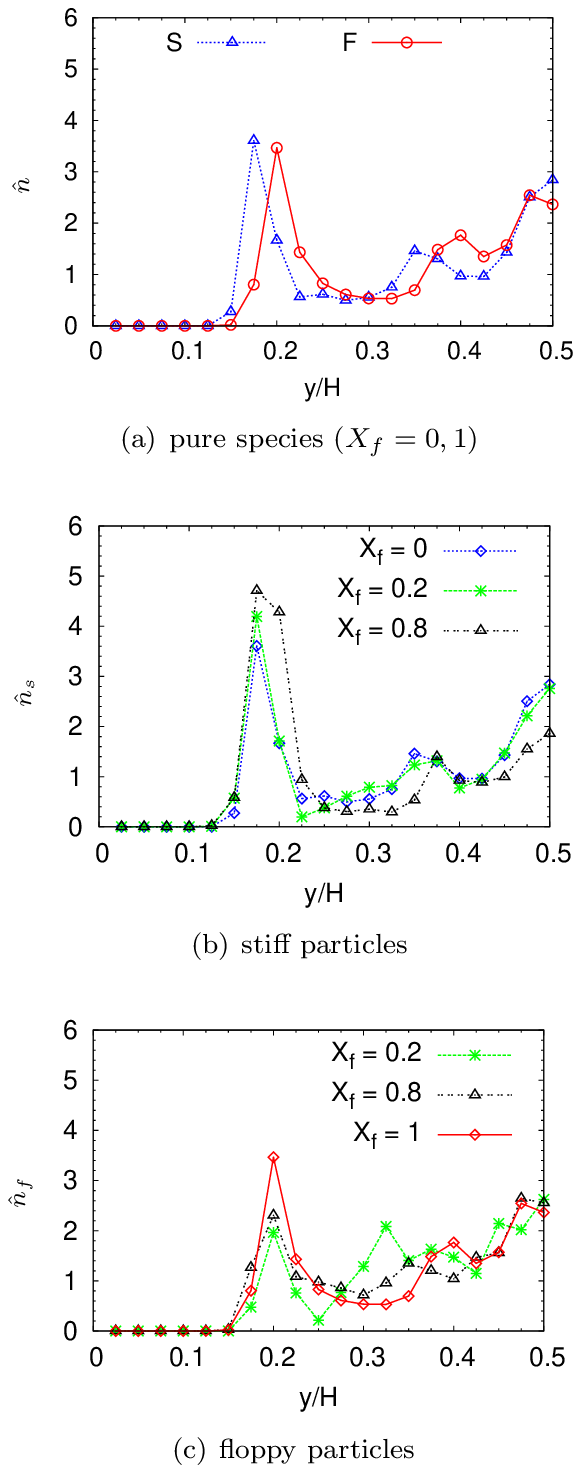}
\caption{(color online) Number density distribution of a given species in the wall normal 
direction normalized by the mean number density of that species, $\hat{n}(y)$.
The channel wall is at $y=0$ and the centerline at $y=0.5H$.
Note that a uniform distribution will yield a value of one throughout. (a) Pure species: $X_f=0$ \& $X_f=1$, 
(b) $\hat{n}_s(y)$ for stiff particles at various values of $X_f$, and (c) $\hat{n}_f(y)$ for floppy  particles at
various values of $X_f$.}\label{fig:n50} 
\end{figure}

\begin{figure}[!t]
\centering
\includegraphics[width=0.35\textwidth]{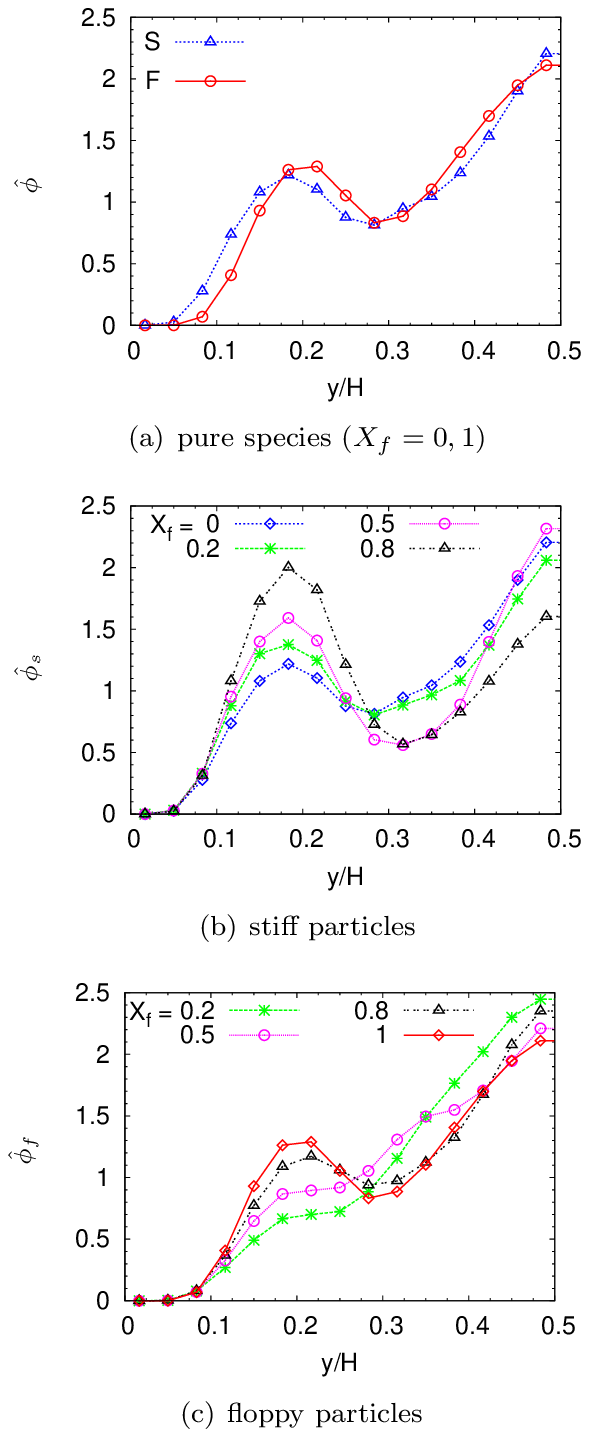}
\caption{(color online) Volume fraction distribution of a given species in the wall 
normal direction normalized by the mean volume fraction of that species, $\hat{\phi}(y)$.
The channel wall is at $y=0$ and the centerline at $y=0.5H$. A uniform distribution will
yield a value of one throughout. (a) Pure species: $X_f=0$ \& $X_f=1$, (b) $\hat{\phi}_s(y)$ for 
 stiff particles at various values of $X_f$, and (c) $\hat{\phi}_f(y)$ for floppy particles
 at various values of $X_f$.}\label{fig:prob50}
\end{figure}

To gain further insights into the particle distribution, Figs. (\ref{fig:n50}) and 
(\ref{fig:prob50}) show $\hat{n}(y)$ and $\hat{\phi}(y)$,  respectively. Note that in these figures, the channel wall is at the left and the centerline at the right -- results in each half of the slit have been combined to enhance the statistics.
We first examine Fig. (\ref{fig:n50}a), which shows $\hat{n}(y)$ for
suspensions  of pure species ($X_f=0$ and $X_f=1$).
In this plot two peaks are very apparent: a sharp peak close to the
wall and a broad peak close to the centerline. Even though the peak
close to the wall looks sharper in comparison to the one at the centerline, the width of this
peak is relatively small; hence the volume fraction around the sharp peak is much
lower in comparison to the volume fraction around the centerline (Fig. \ref{fig:prob50}a).
Also, the peak next to
the wall is closer to it for the stiffer particles than for 
the floppy particles, thereby indicating that the cell free layer
is wider in suspensions of more deformable particles consistent with other simulations \citep{bagchi09}.
The presence of peaks also points toward the formation of layers,
though particle snapshots (see Fig. \ref{fig:50snap}) and 
animations reveal that only the peaks near the walls 
are representative of layers, while much of the interior region
consists of disorganized particles. This is similar to observations
is confined suspensions of drops \citep{pozrikidis2000}. The layering 
close to the walls in concentrated suspensions is easily explained 
by recalling that the wall is a no penetration boundary, thereby promoting
the formation of layers next to it. It is also worth pointing out
that the volume fraction at the centerline (Fig. \ref{fig:prob50}a)
is about twice the average volume fraction. Hence the suspension
is concentrated around the centerline with the volume fraction there being $\phi \approx 0.4$.

We next examine the effect of varying $X_f$ on
$\hat{n}(y)$ and $\hat{\phi}(y)$ of individual species.
We focus first on the stiff particles,
for which $\hat{n}_{s}(y)$ is shown in Fig. (\ref{fig:n50}b) and $\hat{\phi}_{s}(y)$ in Fig. (\ref{fig:prob50}b).
It is clearly seen that with increasing fraction of floppy particles,
the concentration of stiff particles increases in the near wall region;
at low values of $X_f$ this increase is mostly compensated by the decrease 
in concentration in the region between the two peaks (one at the centerline
and one next to the wall), while at higher values of $X_f$, the stiffer
particles are also slightly depleted in the region around the centerline.
For the floppy particles (Fig. \ref{fig:n50}c and \ref{fig:prob50}c),
the exact opposite behavior is observed, namely, with increasing
fraction of stiffer particles (i.e. decreasing $X_f$), floppy particles get depleted
in the region close to the walls and accumulate in the region
between the two peaks; no significant changes in the concentration are observed in the region close to the centerline, except at very low values of $X_f$.

\begin{figure}[!t]
\centering
\includegraphics[width=0.35\textwidth]{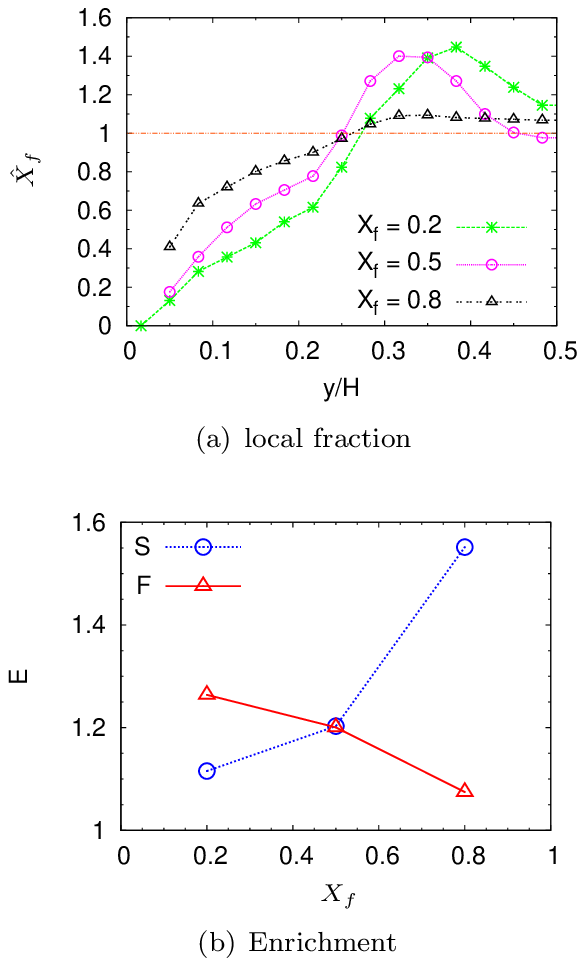}
\caption{(color online) (a) Local fraction of the total particle
volume contributed by floppy particles normalized by its bulk value, $\hat{X}_f$.
(b) The enrichment factor $E$ for both floppy (F) and stiff (S) particles as a function of $X_f$.}\label{fig:50enr}
\end{figure}

We turn now to quantifying the degree of segregation in the mixture. 
Here, we employ a particularly simple measure based on the volume 
fraction distribution of the two species as a function of the wall
normal coordinate. To begin with, we compute $\tilde X_f(y)$ defined as
the fraction of the total \textit{particle volume} contributed by
floppy particles as a function of the wall normal coordinate:
\begin{equation}
\tilde X_f(y) = \frac{\phi_f(y)}{\phi_s(y)+\phi_f(y)}.
\end{equation}
This fraction normalized by the mean number fraction of the floppy particles $X_f$ is
denoted by $\hat{X}_f(y)$: $\hat{X}_f(y) = \tilde X_f(y)/X_f$, which essentially gives the local enrichment 
(when $\hat{X}_f(y)>1$) or depletion  (when $\hat{X}_f(y)<1$) of the floppy particles. 
We show in Fig. (\ref{fig:50enr}a) the plot of $\hat{X}_f$ 
as a function of the wall normal coordinate ($y$) for various values of $X_f$.
As expected, in all cases, floppy particles are enriched in the region close to the 
centerline, while the stiffer particles are enriched in the region close
to the walls. If one's goal is to separate these two species, then an obvious
way to achieve this is by separating the suspension present in these two regions.
In this particular case, the plane separating these two regions is found to be
$y \approx H/4$. We then define an enrichment factor $E$ for each species
as the ratio of its fractional contribution to the total particle volume
in its \textit{enriched} region normalized by the corresponding bulk value. For example
the enrichment factor for floppy particles $E_f$ is obtained as 
\begin{equation}
E_f =  \frac{1}{X_f} \left( \;\frac{\int \phi_f(y) dy}{\int (\phi_s(y) + \phi_f(y))dy} \right),
\end{equation}
where the integrals in the above equation are restricted to the 
region where $\hat{X}_f \geq  1$. We remark that, by definition, $E \geq 1$.
We show the plot of $E$ for both species as a function of the mean 
fraction of the floppy particles  $X_f$ in Fig. (\ref{fig:50enr}b). This result shows
that the enrichment of either species increases as it becomes more
dilute. Moreover, the rate of this increase is much more rapid for
the stiffer particles than for the floppy particles. This slow
increase for the floppy particles is perhaps not too surprising, as its enriched region is 
centered around the centerline, where even the stiffer particles have a 
propensity to accumulate, thereby reducing the relative enhancement of
the floppy particles.

\begin{figure}[!t]
\centering
\includegraphics[width=0.35\textwidth]{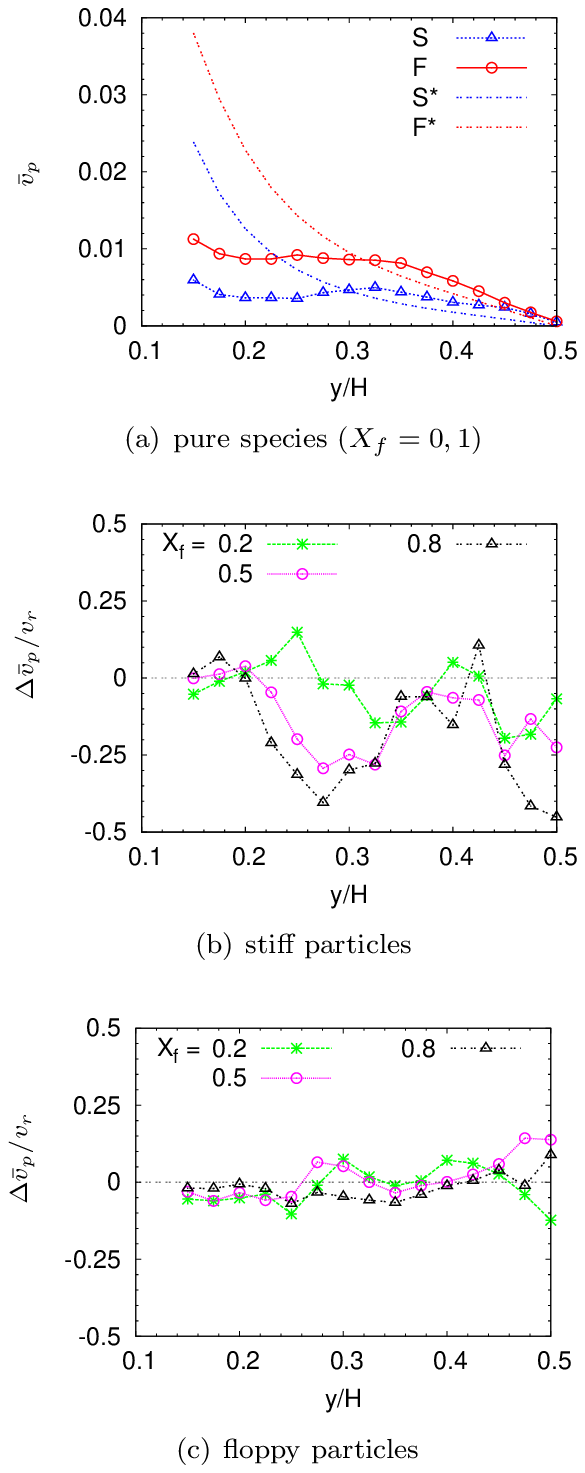}
\caption{(color online) (a) Average velocity \vp of a particle of a given type in suspensions of pure
species (S \& F) as a function of $y$. Also shown are the velocities of an isolated (single) stiff (S*) and floppy (F*) particles in the same geometry and flow. (b) Relative change in the velocity of the stiff particles 
at a given $X_f$ over the corresponding pure species result in (a). (c) Same
as (b), but for the floppy particles.}\label{fig:50vbar}
\end{figure}

\subsection{Particle velocity in the wall normal direction}\label{sec:vp}
So far we have focused on the particle distribution in the wall normal direction.
It will also be of great interest to examine the particle 
velocity in the wall normal direction, as that is
expected to play an important role in its distribution 
in the wall normal direction. We first show, in Fig. (\ref{fig:50vbar}a), the mean wall normal particle
velocity $\bar{v}_p(y)$ in suspensions of pure species ($X_f=0$ and $X_f=1$).
This is obtained by first dividing the channel height into bins. A given particle's instantaneous
wall normal velocity is then assigned to a bin based on the wall
normal coordinate of its center of mass; the average
velocity in a bin then gives the $\bar{v}_p$ in Fig. (\ref{fig:50vbar}a).
For comparison, we also plot in the figure the quasi-steady wall normal 
velocity of an \textit{isolated} capsule.
This quasi-steady velocity is obtained by holding the center
of mass of the isolated particle steady at a given location
-- at the end of each timestep, the particle was translated so that
its center of mass returned to the original position.
The resulting steady state particle velocity is referred to as the isolated
particle quasi-steady migration velocity.

The mean particle velocity $\bar{v}_p$ in the suspension appears
to show three different regimes (Fig. \ref{fig:50vbar}a).
In the region close to the walls ($y/H<0.2$),
$\bar{v}_p$ is a decreasing function of the distance from the walls, much like the
migration velocity of an isolated particle, though the velocity
in the suspension is significantly lower. Recall that $y/H=0.2$ approximately
corresponds to the peak of $\hat{n}(y)$ and $\hat{\phi}(y)$
(Figs. \ref{fig:n50} \& \ref{fig:prob50}). In the intermediate region
($0.2H < y <0.3H$), $\bar{v}_p$ is approximately a constant, and
can even be a slightly increasing function of distance from the wall$y$. In the
region close to the centerline ($0.3H < y <0.5H$), the velocity
decreases linearly with distance. This behavior is similar to that of an isolated
particle, whose migration velocity close to the centerline 
also decreases linearly with distance. This results from the fact the
shear rate decreases linearly with distance in Poiseuille flow,
while the gradient of the shear rate is a constant. The migration
velocity, which, to leading order near the centerline, is the product of these two
terms, therefore decreases linearly with distance \citep{leal80}.
Perhaps the most interesting feature in Fig. (\ref{fig:50vbar}a)
is the fact the mean particle velocity in the suspension can
be \textit{higher} than the corresponding isolated particle migration velocity.
This effect certainly arises due to particle-particle interactions and 
is discussed further below. 

We next turn to the effect of $X_f$ on $\bar{v}_p$ of particles of a given species in the suspension.
To quantify this, we compute the relative change in the velocity $\Delta \bar{v}_p/v_r = (\bar{v}_p-v_r)/v_r$,
where $v_r$ refers to $\bar{v}_p$ in the corresponding pure species suspension. We show
plots of $\Delta \bar{v}_p/v_r$ for the stiff particles at several values of $X_f$  in
Fig. (\ref{fig:50vbar}b), while the same plots for the floppy particles are shown 
in Fig. (\ref{fig:50vbar}c). Focusing first on the stiff particles, we find
that their mean wall normal velocity $\bar{v}_p$ is greatly reduced, in some cases by nearly 
$50\%$, as the fraction of floppy particles in the suspension increases. The 
most significant decrease in the velocity is observed in the second region
discussed above, where $\bar{v}_p$ of the  particles in suspensions of pure species is approximately
a constant (Fig. \ref{fig:50vbar}a). A significant decrease is also observed in
a thin region around the centerline, though,  $\bar{v}_p$ itself is very small here.
To gain some insights into this behavior, we remark that with increasing fraction of the
floppy particles, a stiff particle is expected to ``collide" more frequently with floppy particles
than with particles of the same kind, particularly from its side facing the centerline where most of
the floppy particles are concentrated. Therefore, it appears that the mean additional effect
of these heterogeneous collisions is to push the stiffer particles toward the wall.
Later, in Sec. (\ref{sec:mech}), it will be shown that this observation is 
consistent with simple model studies on homogeneous and heterogeneous pair collisions.
We next consider the effect of $X_f$ on $\bar{v}_p$ of floppy particles (Fig. \ref{fig:50vbar}c).
It is immediately obvious that, in comparison to $\bar{v}_p$ of stiffer particles, $\bar{v}_p$ of 
floppy particles is a weak function of $X_f$ and no particular conclusion can be drawn based on
these trends.

\begin{figure}[!t]
\centering
\includegraphics[width=0.35\textwidth]{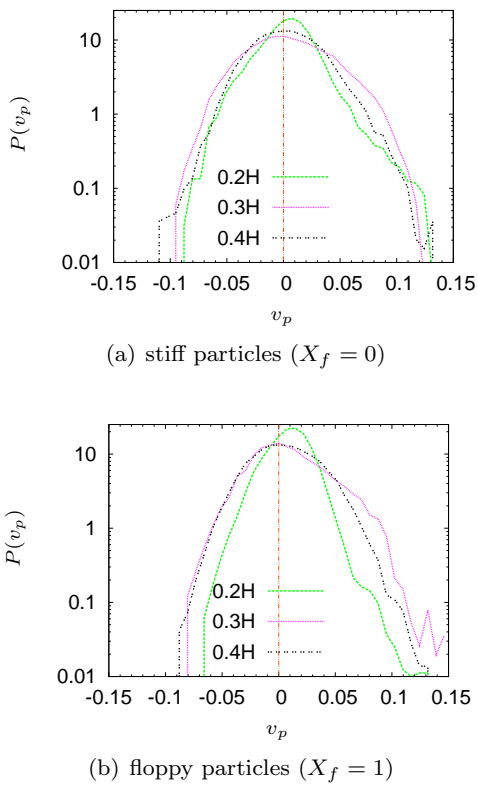}
\caption{(color online) Probability distribution of the wall normal particle velocity $v_p$
at several values of $y/H$. (a) pure stiff particles ($X_f=0$), and (b) pure
floppy particles ($X_f=1$).}\label{fig:50vdist}
\end{figure}

Above, we examined the \textit{mean} particle velocity $\bar{v}_p$.
This particle velocity also exhibits considerable fluctuations about this
mean. To emphasize this, we show in Fig. (\ref{fig:50vdist}) the probability distribution  
of particle velocities $P(v_p)$ on a log-linear scale at various wall normal coordinates
 in suspensions of pure stiff and floppy particles, i.e. $X_f=0$ and $X_f=1$, respectively.
 Not only do these plots confirm
that there are considerable fluctuations in particle velocity (this is quantified below),
but the exact nature of these distributions is interesting in itself and deserves
a brief discussion. Perhaps the most interesting observation in these
plots is that the distribution of positive velocities for particles
close to the wall ($y=0.2H$ in the figure) is non-Gaussian, and 
is, in fact, much closer to an exponential distribution. At larger
values of $y/H$, this changes to a Gaussian distribution. In contrast,
the distribution of negative velocities appears to be Gaussian
at all values of $y/H$. Since one usually expects a Gaussian 
distribution for a variable that is obtained as 
a sum of many uncorrelated parts (in view of the central limit 
theorem \citep{gardiner04}), the appearance of an exponential
distribution perhaps implies contributions from correlated parts \citep{drazer02}.
In the present case, the underlying
physics of the problem provides a basis for these observations. 
We note that a particle is expected to show a non-zero wall normal velocity both
due to ``random" collisions (interactions) with its neighbors, as well as due to the
 non-random effects of the shear rate gradient and hydrodynamic wall repulsion \citep{leal80,leighton91}. In light of the 
above discussion, therefore, the former contribution can be expected to lead to 
a Gaussian distribution, while the latter to a non-Gaussian one. Since both the 
shear rate gradient and wall repulsion lead to positive velocities,
it is therefore not surprising that one observes a non-Gaussian 
positive velocity distribution, especially close to the walls where
the non-random effect is dominant. Negative velocities, on 
the other hand, can result only due to interactions of a particle
with its neighbors, and therefore can be expected to have a Gaussian distribution.

\begin{figure}[!t]
\centering
\includegraphics[width=0.35\textwidth]{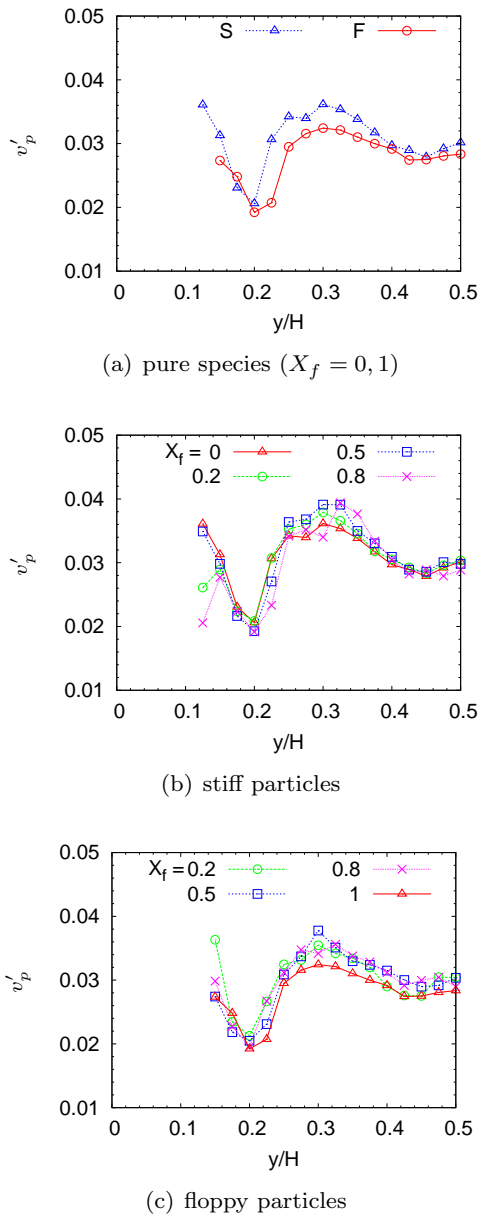}
\caption{(color online) Fluctuation (standard deviation) of the particle velocity of a given species
as a function of $y/H$. (a) In suspensions of pure species ($X_f=0$ \& $X_f=1$),
(b) for stiffer particles at various values of $X_f$, (c) for floppy particles at various values of $X_f$.}\label{fig:50vfluc}
\end{figure}

We next quantify the fluctuation in the particle velocity by computing its
standard deviation, denoted by $v^\prime_p$. We first show in Fig. (\ref{fig:50vfluc}a)
plots of $v^\prime_p$ as a function of $y$ in suspensions of pure species.
As can be seen, the fluctuations in the velocity are much larger than the
corresponding mean, especially in the region close to the centerline.
This is particularly true for the stiff particles as they
not only show a smaller $\bar{v}_p$, but also a higher $v^\prime_p$.
We also find in the plot that the fluctuations show a 
minimum in the region near the wall ($y \approx 0.2H$).
This minimum in fluctuation appears to result from the orderly motion of particles
in the layer formed next to the wall (Fig. \ref{fig:prob50}a).
Moving further away from the wall, we observe a maximum in the
fluctuation $v^\prime_p$ at approximately $y=0.3H$. This position
corresponds well with the minimum in $\hat{\phi}(y)$ in Fig. (\ref{fig:prob50}).
The maximum in the fluctuation implies that a particle at this position undergoes more frequent
collisions with other particles than do particles at neighboring positions, and hence a particle in this region is expected to drift 
away from this region. It is not surprising, therefore, that the position of the maximum in 
velocity fluctuation corresponds well with the minimum in $\hat{\phi}(y)$.
In addition, we observe that the fluctuation amplitude in the region close to the centerline 
is non-zero, and stays finite even at the centerline ($y=0.5H$)
where the local shear rate is zero. This is a well known result in monodisperse suspensions and has 
been attributed to non-local effects due to the finite particle size \citep{nott94,zhao11}. 

We next summarize the effect of $X_f$ on the velocity fluctuations
of the stiffer particles in Fig. (\ref{fig:50vfluc}b), and for the floppy particles
in Fig. (\ref{fig:50vfluc}c). These plots show that the velocity
fluctuation is a weak function of $X_f$, though, in general, the fluctuations
were higher in mixtures than in the corresponding suspensions of pure species.
We also point out that the fluctuations in the velocity of stiffer particles
was always higher than the corresponding fluctuations in the velocity of floppy particles.

Based on the particle velocity fluctuations above, which in turn are rooted in particle collisions,
one may explain the enhancement in $\bar{v}_p$ of a particle in the suspension
over the corresponding isolated particle near the centerline (Fig. \ref{fig:50vbar}a).
Because of the variation in the velocity fluctuations (collisions) as a function of $y$,
one can expect a particle to drift down the gradient of velocity fluctuations (collision) \citep{phillips92,kumar11b};
the velocity fluctuation profile in Fig. (\ref{fig:50vfluc}a) then implies a drift
toward the centerline due to this effect, thereby leading to an enhancement in
the $\bar{v}_p$. 

Results with other simulation parameters in Tb. (\ref{tb:runs})
show a qualitatively similar behavior and are presented in the appendix for completeness. 
In the following section, we explore the mechanism responsible
for the segregation behavior.

\section{Development of mechanistic model}\label{sec:mech}
In this section, we elucidate the mechanism that leads to the segregation
between the stiff and the floppy particles in suspensions of mixtures.
In particular, we seek to explain the following
two observations on the segregation behavior: (i) in a suspension of primarily floppy
particles (i.e. large $X_f$), the stiff particles accumulate in the near wall 
region, while being depleted in the region around the centerline, and (ii)
in a suspension of primarily stiff particles, the floppy particles are 
depleted from the near wall region accompanied by an increased accumulation
in the region near the centerline. In these statements, the degree of 
depletion and excess of a given species in a particular region is with
respect to the distribution of that species in its pure suspension (i.e. $X_f=0$ or $X_f=1$).
We will show that these observations can be qualitatively explained
with a unified mechanism incorporating the wall-induced migration
of deformable suspended particles away from the wall and
the particle fluxes associated with heterogeneous and homogeneous pair collisions.

\subsection{Framework}
To make progress toward elucidating the segregation mechanism, it will be helpful to first 
develop a modeling framework with which to analyze and interpret the simulations results.
In particular, we focus on the particle motion in the wall normal 
direction and identify the distinct mechanisms that 
result in such a motion.

An isolated deformable particle in a pressure driven flow acquires a non-zero 
wall normal migration velocity both due to hydrodynamic interactions with
the walls (wall repulsion) \citep{leighton91} and due to the curvature in the flow profile
(shear rate gradient) \citep{leal80}. In Sec. (\ref{sec:resA}), we presented
the quasi-steady migration velocity of an isolated capsule in Poiseuille flow; see
Fig. (\ref{fig:50vbar}). For the present section, we denote this velocity by $V^S$,
where the superscript `S' signifies the origin as the single (isolated)
particle motion. 

In a suspension of particles, the interparticle interactions
also contribute to the particle's wall normal drift velocity \citep{pozrikidis2000}.
In general, developing exact expressions for the
particle velocity in the suspension is difficult, although for dilute solutions the treatment is
relatively simple. This will prove sufficient for our
present, qualitative purposes. In the dilute limit, as an approximation, we can sum 
velocity contributions resulting from  the isolated particle 
motion and due to contributions from particle interactions treated
in a pairwise fashion. The neglect of three-particle and higher order
interactions will cause an error of $O(\phi^2)$, which is small 
in dilute suspensions. We will denote the contribution to a particle's
velocity resulting from pair collisions by $V^P$, where the 
superscript `P' signifies the origin of the term being pair
collisions. We can combine the contributions from pair collisions 
and single particle motion to obtain the total wall normal 
particle drift velocity as:
\begin{equation}\label{eq:vdt}
V \; = \; V^S + V^P.
\end{equation}
We next develop an expression for the velocity of a particle
resulting due to pair collisions ($V^P$).

The (ensemble-averaged) velocity of a particle arising due to pair collisions in a dilute suspension can be 
expressed by a precise integral expression; see, e.g.,  \citet{pozrikidis2000}.
For our qualitative discussion in the present work, however, we employ and expand on
the much simpler phenomenological model of \citet{phillips92}. 
In this model, the number of collisions experienced by a test 
particle is taken to scale as $\dot{\gamma}\phi$. The variation of this
collision frequency over a distance of $O(a)$ is then given
by $a \nabla(\dot{\gamma}\phi)$, where $\nabla$ here reduces to the derivative
with respect to $y$. Furthermore, each of these collisions
are assumed to cause a displacement of $O(a)$, thereby leading
to a particle velocity given by: 
\begin{equation}\label{eq:vp}
V^P = -K a^2 \nabla (\dot{\gamma}\phi),
\end{equation}
where $K$ is a dimensionless constant that may depend on the properties of the particles in the suspension (e.g.~capillary number, shape). A few words are in order 
about the interpretation of this parameter. We note that the above 
expression was derived under the assumption that the particle undergoes
a displacement of $O(a)$ in pair collisions, which is an appropriate order
of magnitude estimate, though it does not provide precise information
about the actual characteristic displacement in pair collisions. 
This information takes additional significance when considering 
a suspension of mixtures, as the characteristic displacement in pair collisions
is expected to be different when a test particle undergoes collision with a
particle of the same type (homogeneous collision) or with a particle of the 
other type (heterogeneous collision). Furthermore, the characteristic displacement
will also depend on the identity of the test particle, i.e. whether it is stiff 
or floppy. In order to make the above model (Eq. \ref{eq:vp}) cognizant of this
information, we view `$K\,a$' as the characteristic displacement of a particle in
pair collisions. For example, when considering pair collisions between smooth rigid
spheres, we have $K=0$ as the characteristic displacement in such collisions
vanishes owing to the symmetry of the geometry and the Stokes flow reversibility.
With this aspect clarified, we generalize Eq. (\ref{eq:vp}) for a suspension
of mixtures to obtain:
\begin{subequations}\label{eq:vsf}
\begin{equation}
V_s^P = -a^2  \left(K_{ss}\nabla(\dot{\gamma}\phi_s)+K_{sf}\nabla \, (\dot{\gamma}\phi_f)\right),
\end{equation}
\begin{equation}
V_f^P = -a^2  (K_{fs}\nabla(\dot{\gamma}\phi_s) +K_{ff}\nabla(\dot{\gamma}\phi_f)),
\end{equation}
\end{subequations}
where the subscript `s' and `f' denote the stiff and the floppy particle, respectively.
We next discuss results from several pair collision studies as 
they provide important insights into the relative values of the characteristic displacements
in various scenarios, i.e. it provides information about the relative values 
of the various constants $K$ in equation (\ref{eq:vsf}).

\begin{figure}[!t]
\centering
\includegraphics[width=0.3\textwidth]{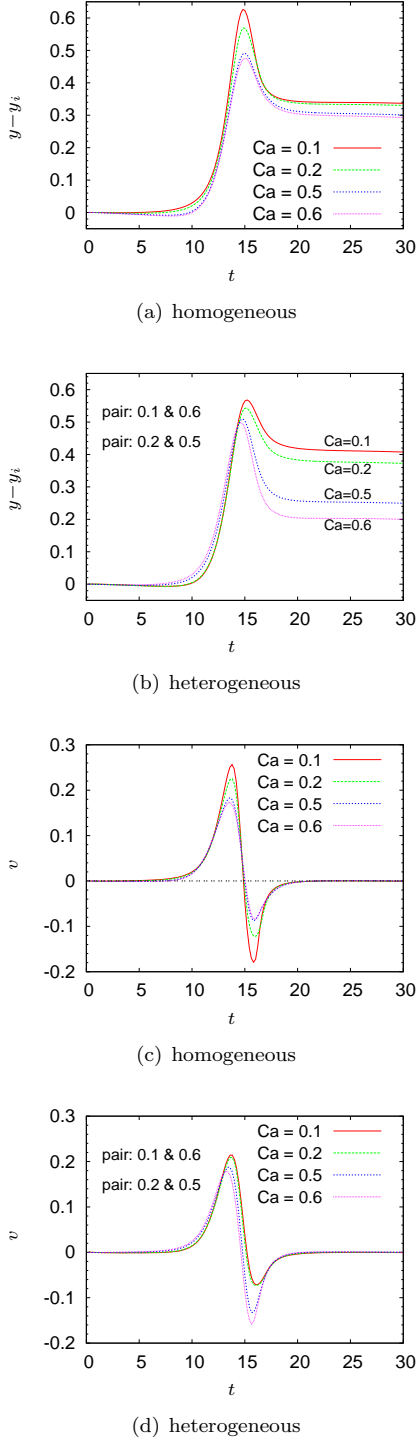}
\caption{(color online) Pair collisions: Displacement of a particle relative to its initial position ($y-y_i$) with time 
in (a) homogeneous pair collisions and (b) heterogeneous pair collisions.
Wall normal velocity of a particle as a function of time in (c) homogeneous
pair collisions and (d) heterogeneous pair collisions.}\label{fig:pair_coll}
\end{figure}

\subsection{Pair collisions}\label{sec:pair}
We presented above the particle velocity resulting from pair collisions between particles,
which obviously plays an important role in the particle distribution along the wall
normal direction. To gain more insights into this process, it will be helpful to examine 
some pair collision results, particularly for pair collisions between two different species, as
no prior study appears to have addressed  this problem. 
We begin by describing the problem setup for pair collision studies.
In all studies, a large cubic simulation box of side $30a$ was considered,
and the particles were placed on either side of the central plane with a given initial separation.
Further, in all these studies, only simple shear flows were considered. 
These choices for the problem setup minimize the single particle migration effect, and hence
provide insights into the effect of particle-particle interactions (collisions).

We begin by considering pair collisions between the same  species (homogeneous collisions)
and show in Fig. (\ref{fig:pair_coll}a) the displacement of the `top' particle in the wall normal direction relative to its initial position ($y-y_i$)
as a function of time. For this study and those below,
we set the initial separation between the two particles in the wall normal direction 
as $\Delta y_i=0.5a$. Other studies with $\Delta y_i =0.25a$ and
$\Delta y_i = 0.75 a$ gave similar trends. In the above figure, it is clearly seen that in this range of 
$\Ca$ ($0.1 \leq \Ca \leq 0.6$), the final displacement ($\delta_{\infty} = y_{\infty}-y_i$)
increases with decreasing $\Ca$. We remark that $\delta_{\infty}$ is known to be
a non-monotonic function of $\Ca$, such that at low enough enough $\Ca$, $\delta_{\infty}$
decreases with decreasing $\Ca$ \citep{pratik10}.

We next consider collisions between two particles with different $\Ca$ (Fig. \ref{fig:pair_coll}b).
From these results we see that the stiffer particle undergoes
much larger displacement ($\delta_{\infty}$) than the corresponding floppy particle involved
in the collision. In addition, the difference in the final displacements of the two species
is larger for larger rigidity ratio system. Further insights are obtained 
by comparing the instantaneous wall normal velocities of the particles during the collision event
for the  cases in Figs. (\ref{fig:pair_coll}a)  and (\ref{fig:pair_coll}b),
which are shown in Figs. (\ref{fig:pair_coll}c) and (\ref{fig:pair_coll}d), respectively.
Focusing first on the result for particle velocity in homogeneous pair collisions (Fig. \ref{fig:pair_coll}c),
we find the particle velocity is positive in the approach part of the collision trajectory,
while it is negative in the receding part of the collision trajectory. As a result of this feature,
a fraction of the maximum displacement in Fig. (\ref{fig:pair_coll}a) is 
recovered in the receding part of the collision trajectory. It is obvious from Fig. (\ref{fig:pair_coll}c)
that the magnitude of velocity in either part of the trajectory is higher 
for the stiffer particle than the floppy particle in homogeneous collisions.
In contrast, in heterogeneous collisions (Fig. \ref{fig:pair_coll}d), while
the magnitude of the velocity is higher for the stiff particle in the approach
part of the trajectory, in the receding part it is the floppy particle which 
exhibits a greater magnitude of velocity. This reversal in the trends of
the velocity in the receding part of the trajectory leads to the observed 
behavior in heterogeneous pair collisions. Summarizing the results of 
this section, we have the following relationships between
the various constants $K$ in Eq. (\ref{eq:vsf}):
(i) $K_{sf} > K_{fs}$, (ii) $K_{sf} > K_{ss}$, and (iii) $K_{fs} < K_{ff}$.
We next discuss the implications of these pair collision results on the segregation behavior.

\begin{figure}[!t]
\centering
\includegraphics[width=0.3\textwidth]{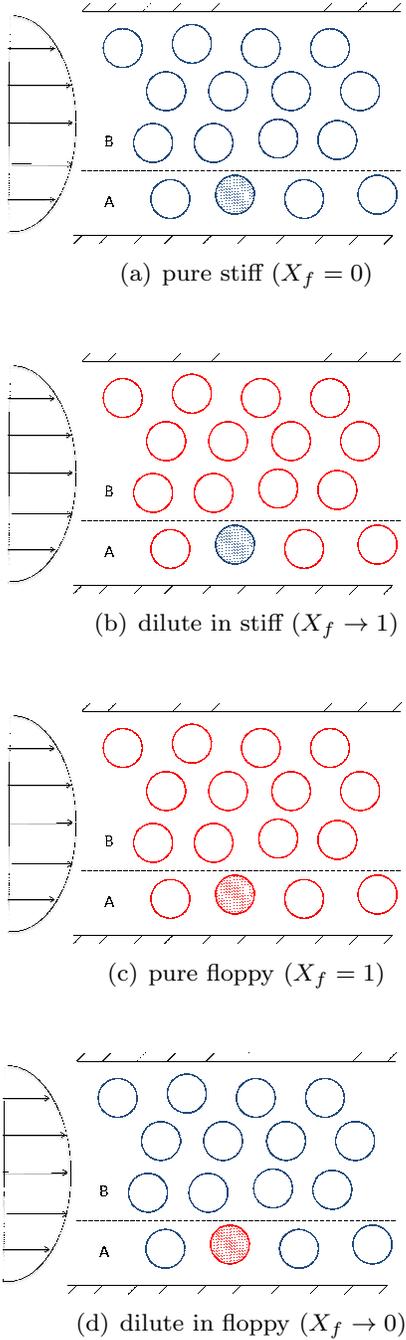}
\caption{(color online) A cartoon depicting the characteristics of the neighborhood of 
a particle of given type (shaded) in the near wall region at various
values of $X_f$. The stiff particle is shown light (blue online), while the floppy particle is
	shown dark (red online).}\label{fig:cartoon_mech}
\end{figure}

\subsection{Mechanism of flow induced segregation}
Based on the results of pair collision studies and the single particle migration velocity,
we are now in a position to identify the mechanism responsible for the segregation 
behavior. In a confined suspension, the wall plays an important role as it
breaks the translational invariance in the wall normal direction. So, we 
divide the overall system domain in two parts: first is the near wall region,
and second is the interior region. These two regions are pictorially
shown in Fig.  (\ref{fig:cartoon_mech}), where
the near wall region is labeled as `A', while the interior region
is labeled as `B'. We first focus on the near
wall region as that turns out to be critical to the segregation behavior.
For a particle in the near wall region A (e.g., shaded particles in Fig. 
\ref{fig:cartoon_mech}), the net of effect of pair collisions
is to push to particle toward the wall, as $-\nabla (\dot{\gamma}\phi)$
is negative in this region. In other words, a particle in this region 
undergoes collisions with other particles present only on its upper half and 
consequently gets pushed toward the wall. Therefore, the net effect of these 
pair collisions is to prevent the `escape' of the particles in the near 
wall region A to the interior region B. In contrast, the single particle 
velocity $V^S$, which is always toward the centerline,  aids the particle 
in `escaping' to the interior region B. Since the single particle migration
velocity $V^S$ depends only on the characteristics of the particle, it is 
\textit{independent} of $X_f$. On the other hand, $V^P$, the velocity arising due
to pair collisions, depends strongly on $X_f$ in view of the results of
homogeneous and heterogeneous pair collisions discussed above, i.e. due to 
differences in values of $K_{ss}$, $K_{ff}$, $K_{sf}$ and $K_{fs}$. Therefore,
the frequency with which a particle present in the region A will escape to the
region B will depend on $X_f$ primarily due to the dependence of $V^P$ on $X_f$. 

To aid in visualizing the problem at various values of $X_f$, we consider four
cases as shown in Fig. (\ref{fig:cartoon_mech}): (a) suspension of stiff particles only ($X_f=0$),
	(b) suspension that is dilute in stiff particles ($X_f \rightarrow 1$),
	(c) suspension of floppy particles only ($X_f=1$),
(d)  suspension that is dilute in floppy particles ($X_f \rightarrow 0$).
In the first two cases, our focus is on the behavior
of the stiff particle, while in the last two cases, our focus is
on the behavior of the floppy particle; in all cases the particle
of interest is shaded. Focusing first on a stiff
particle, its velocity $V^P$ in cases (a) and (b) is given by 
$V^P = -K_{ss}a^2 \nabla (\dot{\gamma}\phi)$ and 
$V^P = -K_{sf}a^2 \nabla (\dot{\gamma}\phi)$, respectively.
In Sec. (\ref{sec:pair}), we concluded that $K_{ss} < K_{sf}$,
and hence that the stiff particle in case (a) faces a smaller 
barrier in escaping to region B than the stiff particle 
in case (b). Therefore, we conclude that a stiff
particle gets increasingly localized in the near wall region
as the fraction of floppy particles in the suspension increases.
We next turn our attention to the behavior of the floppy particle and 
consider cases (c) and (d) in Fig. (\ref{fig:cartoon_mech}). The
velocity of the floppy particle in these cases is given
by $V^P = -K_{ff}a^2 \nabla (\dot{\gamma}\phi)$ and 
$V^P = -K_{fs}a^2 \nabla (\dot{\gamma}\phi)$, respectively.
The results of Sec. (\ref{sec:pair}) indicate that $K_{ff} > K_{fs}$,
and consequently that the frequency of escape of the floppy particle 
to region B from region A should increase as the fraction 
of stiff particles in the suspension increases.

\begin{figure}[!t]
\centering
\includegraphics[width=0.35\textwidth]{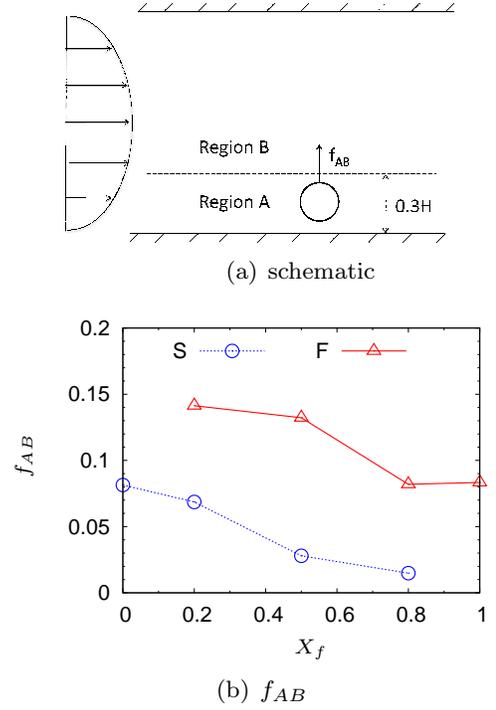}
\caption{(color online) (a) A schematic defining the near wall region A and the interior region B.
	(b) Transition probability $f_{AB}$  defined as the fraction of particles starting
	in region A crossing over to region B in \tm{100}. $f_{AB}$ is shown 
	as function of $X_f$ for both the stiff and the floppy particles.}\label{fig:flux}
\end{figure}

To verify the mechanism presented above, we compute the  probability of
transition of a particle of a given type from region A to region B as a function
of $X_f$ (Fig.  \ref{fig:flux}b). More precisely, this transition probability, denoted by $f_{AB}$,
gives the fraction of particles that are present in region A at a given time, but
cross over to region B within 100 time units from that time.
We have defined the near wall region A as the region extending to a distance of $0.3H$ from either walls, while the remainder of the domain is defined as region B (Fig. \ref{fig:flux}a). The choice of $0.3H$ comes from the minimum of 
$\hat{\phi}$ in Fig. (\ref{fig:prob50}a) and $\hat{n}$ in Fig. (\ref{fig:n50}a),
which essentially separates the near wall region and the interior region.
The transition probability $f_{AB}$ in
Fig. (\ref{fig:flux}b) clearly clearly shows trends consistent with 
the mechanism described above. We focus first on the trends for stiff particles.
At $X_f=0$, we find $f_{AB}^s \approx 0.081$, which gradually decreases 
to $f_{AB}^s \approx 0.015$ at $X_f=0.8$, which amounts to a factor of $5.4$
reduction in the transition probability. This result clearly suggests
that stiff particles get localized in the near wall region as the
fraction of floppy particles increases, which must be due to the 
effect of pair collisions, as the isolated particle migration
velocity is independent of $X_f$. As discussed earlier, this
is the expected trend based on the relative values of $K_{ss}$
and $K_{sf}$, as $K_{sf} > K_{ss}$. We next focus on the 
trends of $f_{AB}$ for the floppy particle in Fig. (\ref{fig:flux}b).
At $X_f=1$, we find $f_{AB}^s \approx 0.083$, which gradually increases
to $f_{AB}^s \approx 0.1413$ at $X_f=0.2$, which amounts to a factor of $1.7$
increase in the transition probability. This implies that,
with increasing fraction of the stiff particles, the floppy
particles escape to the interior region more easily. This
again is expected based on the results of pair collisions
as $K_{fs} < K_{ff}$. 

It is also interesting to note that the $f_{AB}$ for both 
the stiff and the floppy particles 
are very close in their corresponding suspensions of pure 
species, i.e. $X_f=0$ and $X_f=1$, respectively. Results 
of pair collisions above indicate that the $K_{ss}$ and $K_{ff}$
are about the same, though the single particle migration 
velocity $V^S$ is larger for the floppy particle. A larger 
$V^S$ for the floppy particle, in principle, should lead to a
higher $f_{AB}$ for that particle. On
the other hand, a larger $V^S$ may lead to more frequent
collisions with a relatively smaller $\Delta y_i$, the initial separation between
the two colliding particles in the wall normal direction,
as this velocity will cause two particles to approach
each other in the wall normal direction. Since collisions
with smaller $\Delta y_i$ lead to larger cross 
stream displacement $\delta_{\infty}(\Delta y_i)$ \citep{lac07}, the effect of
larger $V^S$ may be nullified by the effect of larger $\delta_{\infty}(\Delta y_i)$. This 
effect enters our model through the effect of $\delta_{\infty}(\Delta y_i)$ on the constant $K$,
as the constant $K$ is essentially the weighted average of $\delta_{\infty}(\Delta y_i)$,
where the weight is the probability for a collision with a given $\Delta y_i$
to occur. Computing this weighted average rigorously is beyond the scope
of the present work.

In addition to the mechanism outlined above, we must also address the 
effects of excluded volume, which is only expected
to amplify the effects of this mechanism.
This is because if the floppy particles find it easier 
to escape to the interior than the stiff particles,
then their enhanced presence in the interior will only make 
the escape of the stiff particles to the interior more
difficult, as only a limited volume fraction of particles can be present
in the interior. So far we have been treating the motion of 
stiff particles and floppy particles separately. The 
excluded volume effect essentially couples the motion
of these two species, which is intuitively expected
to be the case in a suspension of particles, especially at non-dilute volume fractions.

So far we have focused entirely on the near wall region. We now
briefly discuss the dynamics in the interior region. In this
region, a particle has near neighbors both in its upper and
lower half, and hence the gradient in the collision frequency
$a\nabla(\dot{\gamma}\phi)$ is expected to be small relative
to the collision frequency $\dot{\gamma}\phi$ itself  (the
latter collision frequency is only an approximation as it
does not account for finite particle size; see Sec. \ref{sec:vp}).
When coupled with the fact that the single particle velocity
is also small far away from the walls (Sec. \ref{sec:vp}), we
conclude that the particle motion in this region is 
expected to be dominated by shear-induced diffusion.
Therefore, the characteristics of the particle motion in this region are not
expected to play a dominant role in the segregation between the two species.
Nonetheless, the diffusive behavior in this region is necessary for
transporting the particles from the bulk to the near wall region, and 
hence indirectly plays a role in the observed segregation between the two species.

We conclude this section by addressing an interesting feature
concerning the time evolution of the mean absolute distance \dbar 
of a species in a mixture. In Sec. (\ref{sec:dist}),
we observed that in a mixture of stiff and floppy particles, \dbar 
of both species initially decreases with time (Fig. \ref{fig:ycm_t}c).
But, beyond approximately \tm{600} in Fig. (\ref{fig:ycm_t}c),
the \dbar of stiff particles was found to gradually increase 
with time. This phase of the process was termed as a reorganization phase 
between the stiff and the floppy particles, as \dbar of the overall 
suspension was essentially steady in this phase of the simulation.
Based on the mechanism outlined above, this reorganization phase 
can be associated with the transport of the stiff particles in the 
interior region to the near wall region, where they are expected to become nearly irreversibly
trapped.  
Therefore, the time scale after which this reorganization phase is noticeable
in simulations can be associated with the time required for a particle
near the centerline to diffuse to the near wall region.
We therefore see that the mechanism outlined here is capable
of addressing even the finer aspects of the segregation behavior.

\section{Conclusions}\label{sec:conc}
In this article we investigated the segregation behavior in suspensions
of binary mixtures of Neo-Hookean capsules subjected to pressure driven
flow in a planar slit. The two species in the binary mixture
have different membrane rigidities as characterized by their separate
capillary numbers. Detailed boundary integral simulations revealed that with increasing
fraction of the floppy particles in the suspension (i.e.~with increasing $X_f$), 
there is an enhancement in the concentration of the stiff particles in the near
wall region. On the other hand, as the fraction of stiff particles in the suspension
increases (i.e. $X_f$ decreases), a depletion in the concentration 
of floppy particles in the near wall region is observed. 
As a result of this behavior, stiff particles segregate
in the near wall region in a suspension of primarily
floppy particles, while floppy particles segregate around the 
centerline in a suspension of primarily stiff particles.
A novel and detailed mechanism based on the degree of localization
of a particle of a given type in the near wall region as a
function of $X_f$ was developed. The three main ingredients
of this mechanism which governs the localization of a particle
in the near wall region are: (i) the single particle migration velocity
toward the centerline, which provides an avenue for the 
particle to escape to the interior,  (ii) the 
repeated pair collisions of a particle in the near wall 
region with its neighboring particles, which pushes it toward
the wall, thereby preventing its escape to the interior and (iii) the difference in final displacement between homogeneous and heterogeneous pair collisions.
The single particle migration velocity is independent 
of $X_f$, while the cross-stream displacement depends
strongly on $X_f$. In particular, results of heterogenous pair
collisions involving a stiff and a floppy particle showed that
the stiff particle experiences a substantially larger cross-stream displacement
as a result of the collision than the floppy particle.
This result indicates, consistent with simulation results, that as the fraction of floppy particles
in the suspension increases, a stiff particle in the near wall region will get
increasingly localized there due to the strong effect of heterogeneous
pair collisions on its wall-normal displacement. On the other hand,
as the fraction of stiff particles in the suspension increases,
a floppy particle in the near wall region can escape to the interior
much more easily due to the much weaker effect of heterogenous pair
collisions on its wall-normal displacement.
This mechanism, therefore, consistently explains the features of the segregation
behavior observed in detailed numerical simulations. Important topics for future
work will be the extent to which the same picture can explain segregation by size
and shape as well as segregation in the (nominal) absence of shear rate gradients,
as in plane Couette flow. It will also be of interest to formalize the qualitative
picture into a stochastic process model that more quantitatively captures the
simulation and experimental results.


\begin{acknowledgments}
The authors acknowledge helpful discussions with Kushal Sinha, Rafael Henriquez, Pratik Pranay and Juan Hernandez-Ortiz.  This work is supported by NSF grants CBET-0852976 (funded under the American Recovery and Reinvestment Act of 2009) and CBET-1132579.
\end{acknowledgments}

\appendix*

\begin{figure}[!t]
\centering
\includegraphics[width=0.3\textwidth]{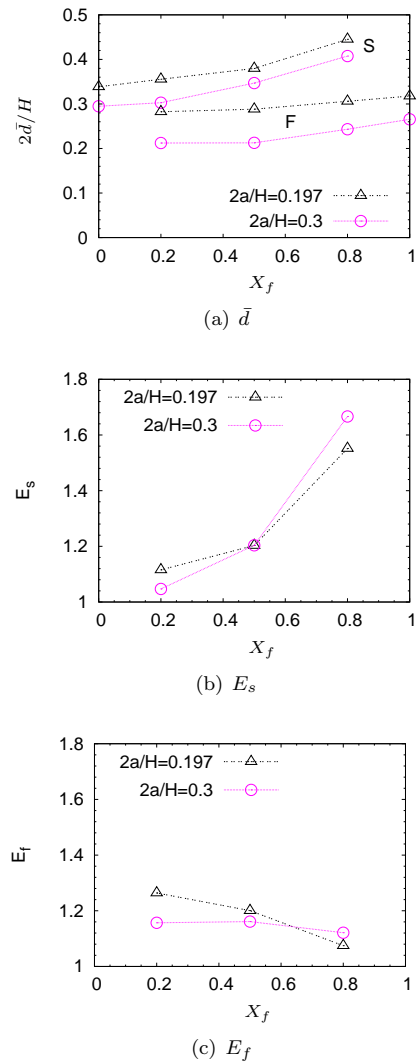}
\caption{(color online) (a) Steady state \dbar as a function of $X_f$ for stiff and floppy
particles at two different confinement ratios. Enrichment factor as a function of $X_f$  for
 (b) stiff particles, and (c) floppy particles.}\label{fig:conf_ycm}
\end{figure}

\begin{figure}[!t]
\centering
\includegraphics[width=0.3\textwidth]{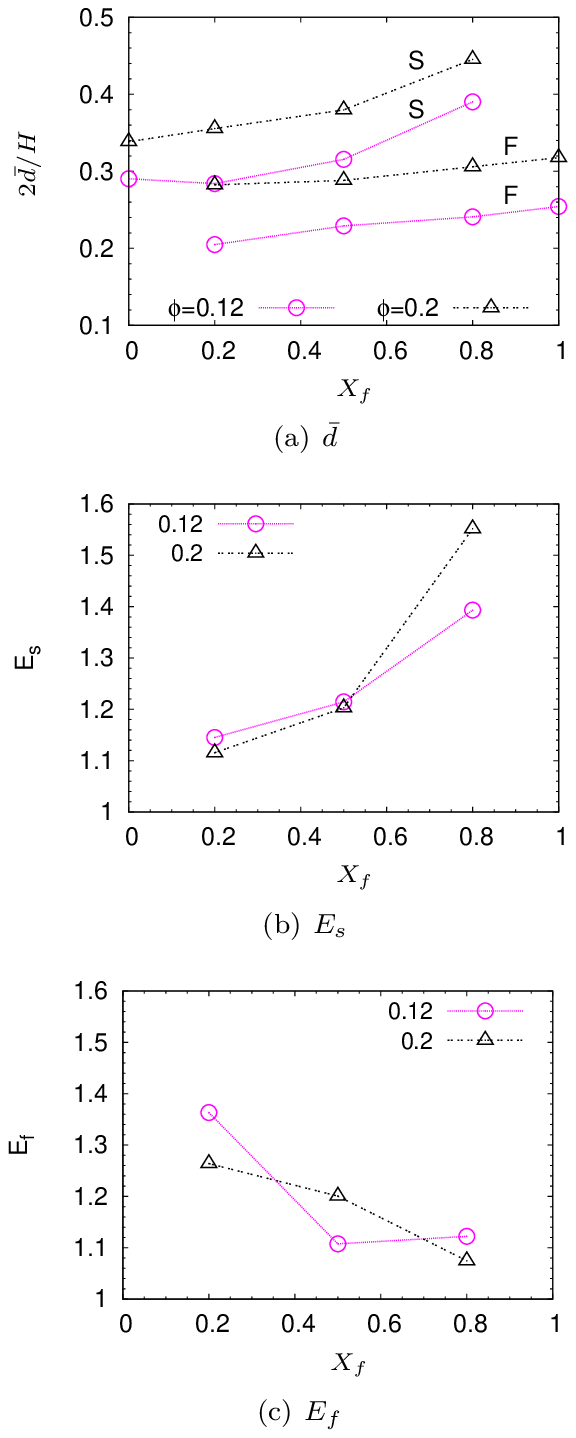}
\caption{(color online) (a) Steady state \dbar as a function of $X_f$ for stiff and floppy
particles at two different volume fractions. Enrichment factor as a function of $X_f$ 
for (b) stiff particles, and (c) floppy particles.}\label{fig:phi_ycm}
\end{figure}

\begin{figure}[!t]
\centering
\includegraphics[width=0.3\textwidth]{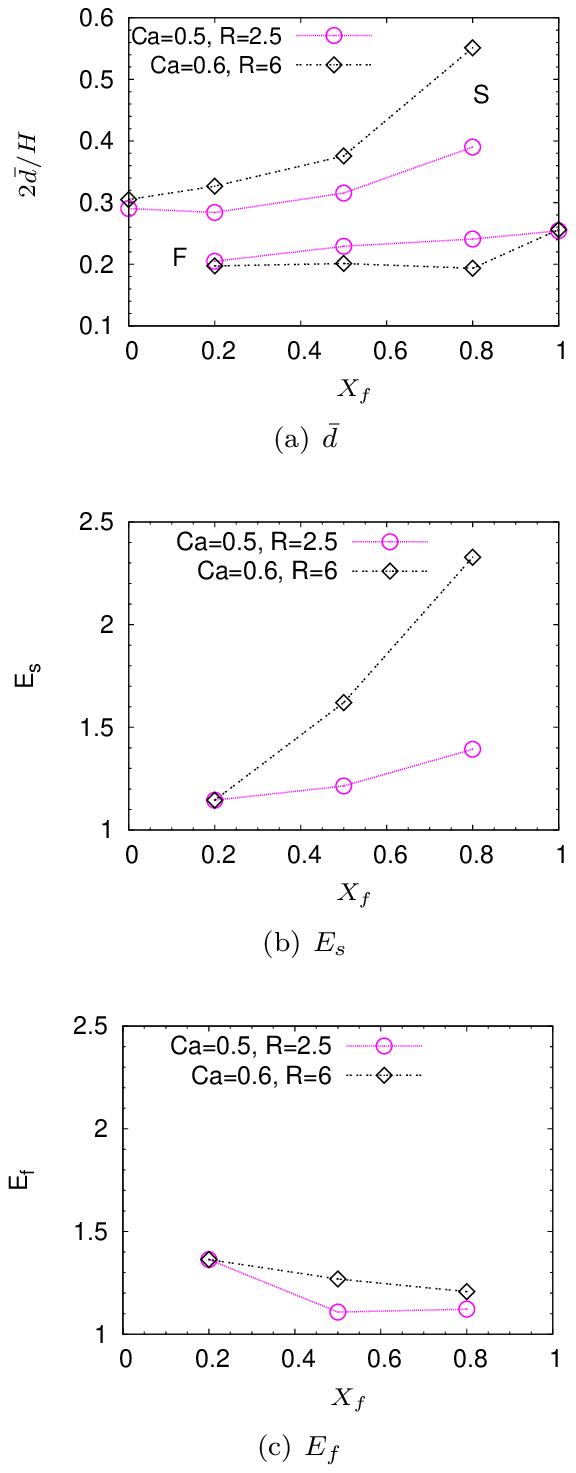}
\caption{(color online) (a) Steady state \dbar as a function of $X_f$ for stiff and floppy
	particles as noted in the figure key. Enrichment factor as a function of $X_f$ for (b) stiff 
	particles, and (c) floppy particles.}\label{fig:R_ycm}
\end{figure}

\section{Effect of confinement ratio, volume fraction and rigidity ratio}
We present here some additional results of simulations with different confinement ratio, volume fraction or rigidity ratio than were used in the main text. The results are all consistent with those results and are included here for completeness.


To examine the effect of confinement ratio on the results,  we consider suspensions with
a volume fraction of \ph{0.2} and two different confinement ratios:
 $2a/H=0.197$ and $2a/H=0.3$. The rest of the parameters for these two simulation sets
 are listed in Tb (\ref{tb:runs}) as sets A and B.
We first show the effect of confinement ratio on \dbar for both the
stiff and the floppy particles in Fig. (\ref{fig:conf_ycm}a) as a function of $X_f$.
As is obvious, \dbar of both the species decreases with increasing confinement ratio, i.e. 
an increasing fraction of the particles accumulate near the centerline with increasing $2a/H$.
Moreover, the trend in \dbar of either species with varying $X_f$ is similar in both cases -- with increasing
fraction of floppy particles (increasing $X_f$), the stiffer particle gets 
increasingly displaced toward the wall, while with increasing fraction
of stiffer particles (decreasing $X_f$), the floppy particles get increasingly 
displaced toward the centerline.

For a more quantitative measure of the segregation behavior, we compute the 
enrichment $E_s$ and $E_f$ of the two species as a function of $X_f$ (Figs. \ref{fig:conf_ycm}b, \ref{fig:conf_ycm}c).
These plots show that $E_s$ increases more rapidly with $X_f$
in more confined suspension ($2a/H=0.3$), while $E_f$ increases more rapidly with
decreasing $X_f$ in less confined suspension ($2a/H=0.197$). In other words, based on this measure,
more confined suspensions lead to a higher maximum segregation of the stiff particles,
while less confined suspensions leads to higher maximum segregation of floppy particles.


We now we briefly examine the effect of volume fraction on the segregation
behavior in the suspension. Results will be reported
here for a suspension with a volume fraction of $\phi=0.12$ and compared
with the results of $\phi=0.2$ suspension discussed above in Sec. (\ref{sec:resA}).
The parameters for this simulation set are listed in Tb. (\ref{tb:runs}) in the
row labeled ``C", which are the same as in set A except for the volume fraction.

We begin by showing \dbar in Fig. (\ref{fig:phi_ycm}a) for both the 
stiff (\ca{0.2}) and the floppy (\ca{0.5}) particles in the suspension as a function of $X_f$.
For comparison, we also show \dbar for the $\phi=0.2$ suspension.
At any given volume fraction, \dbar for stiffer particles
is found to be higher than for floppy particles. 
Moreover, as the fraction of the floppy particles $X_f$
increases, \dbar for stiffer particles increases,
i.e. they get increasingly displaced toward the wall.
On the other hand, as the fraction of stiff particles in the 
suspension increases ($X_f$ decreases), \dbar for floppy 
particles decreases, i.e. they get displaced toward the centerline.
All these trends in \ph{0.12} suspensions are similar to the trends 
in \ph{0.2} suspensions. It is also obvious in the plot that at any given value of $X_f$,
the \dbar for a species increases with increasing $\phi$.
This is an expected result, as at  higher volume fractions the increased 
particle-particle interactions reduces the fraction of particles accumulating
near the centerline, thereby leading to an increase in \dbar.

To quantify the degree of segregation, we compute the enrichment $E$ 
for both the species as a function of $X_f$ (Figs. \ref{fig:phi_ycm}b,\ref{fig:phi_ycm}c).
These plots confirm that the  maximum enrichment of a given species is observed when that species is
present in  dilute amounts. Furthermore, the maximum enrichment
of the stiff particles is higher for the \ph{0.2} suspension, while 
the maximum enrichment of floppy particles is higher for the \ph{0.12} 
suspension.



Finally, we consider the effect of $\Ca$ as well as that of the rigidity ratio $R$ 
on the segregation behavior. For this section, we compare
the results of set C and set D in Tb. (\ref{tb:runs}) --
the volume fraction in both the simulations were \ph{0.12}, while the
stiff particles have a $\Ca$ of $0.2$ and $0.1$, and the floppy particles
have a $\Ca$ of $0.5$ and $0.6$, respectively. These sets of $\Ca$ for the
stiff and the floppy particles give a rigidity ratio of $R=2.5$ and $R=6$, respectively. 

We begin by showing \dbar for both the stiff and floppy particles
in these simulations in Fig. (\ref{fig:R_ycm}a). One can observe in the plot that
\dbar for floppy particles is approximately the same in these two simulations.
In contrast, \dbar for the stiffer particles in the two
cases show a marked difference as a function of $X_f$ -- at higher values
of $R$, the stiffer particle gets increasingly displaced toward the walls
as $X_f$ increases. For quantifying the degree of segregation
between the two species, we again compute the enrichment factor $E$
for both the species as function of $X_f$ (Figs. \ref{fig:R_ycm}b, \ref{fig:R_ycm}c).
As can be observed, $E_s$ shows a much larger increase with increasing $X_f$
for $R=6$ case in comparison to the $R=2.5$ case. In contrast,
the enrichment of the floppy particles, $E_f$, is found to be
nearly the same in both cases. These results suggest that
the degree of segregation of the stiffer particle increases
as the rigidity ratio increases, particularly at higher values of $X_f$.



\end{document}